\documentstyle[11pt,newpasp,twoside,epsf]{article}
\def\edcomment#1{\iffalse\marginpar{\raggedright\sl#1\/}\else\relax\fi}
\reversemarginpar

\begin{document}
\title{Integral Field Spectroscopy with the Gemini 8-m Telescopes}
\author{Bryan W. Miller, James Turner}
\affil{Gemini Observatory, Casilla 603, La Serena, Chile}
\author{Marianne Takamiya, Doug Simons}
\affil{Gemini Observatory, 670 N. A'ohoku Place, Hilo, HI, 97620, USA}

\author{Isobel Hook}
\affil{UK Gemini Project Office, University of Oxford, UK}

\begin{abstract}
We give an overview of the current and future IFU capabilities on the
Gemini 8-m telescopes.  The telescopes are well-suited to integral
field spectroscopy and both telescopes will have optical and
near-infrared IFUs within the next few years.  Commissioning for the 
GMOS IFU on Gemini North has begun recently and it is now available to the
community. Future integral field instruments will take advantage of 
wide-field adaptive optics systems. 
\end{abstract}

\section{Introduction}

The high image quality and large collecting areas of the Gemini 8-m
telescopes make them well-suited to integral field spectroscopy.  On
smaller telescopes signal-to-noise considerations have forced
integral-field units (IFUs) to have either coarse spatial sampling
(e.g. INTEGRAL, DensePak, SAURON) or fine sampling of the highest
surface brightness targets (e.g. OASIS, TEIFU).  More collecting area,
a smaller diffraction limit, and new technology in today's 8-10 meter
telescopes means that higher signal-to-noise spectra of finer spatial
structures can be obtained.  Therefore, the first generation of
instrumentation on Gemini will include optical and near-infrared IFUs
on both telescopes: GMOS and NIFS instruments on Gemini North, and
GMOS and GNIRS on Gemini South (see Table~1).  This paper summarizes
the capabilities of the telescopes and these instruments.

While they are sensitive from optical through mid-infrared
wavelengths, the Gemini telescopes are optimized for observing in the
near and mid-infrared.  Special features of the telescopes include
minimizing the mass above the primary mirror in order to lower the
thermal background and improve airflow over the mirror, future
low-emissivity mirror coatings, daytime climate control, and a
tip/tilt chopping secondary.  In addition, both telescopes will have
facility adaptive optics units that will be able to feed a
near-diffraction-limited beam to any instrument.  Therefore, Gemini
will be able to deliver image quality in the near-infrared about a
factor of two better than HST (Figure~1).  The Gemini IFUs have been
designed to work with the expected image quality from either tip/tilt
or adaptive-optics corrected beams.  The rest of this paper briefly
describes the capabilities of each Gemini integral field unit in
approximate order of when it will come into service.

\begin{center}
Table~1. Summary of Gemini Integral Field Instruments
\begin{tabular}{|l|l|l|l|l|}
\hline
Instrument/ & FOV & Sampling & Wavelength & R \\ 
Location    &     &          & Range      & \\
\hline
GMOS        & $7''\times5'' +$ & $0.2''$ & 0.4--1.1 $\mu$m & 500--8000\\
GN and GS   & $3\farcs5\times5''$ &  & & \\
\hline
GNIRS       & $3\farcs2\times4\farcs4$ & $0.15''$ & 1--5 $\mu$m & 667,2000,\\
GS          &  & & & 6000\\
\hline
NIFS        & $3''\times3''$ & $0.1''$ & 0.9--2.5 $\mu$m & $\sim5000$\\
GN          & & & & \\
\hline
\end{tabular}
\end{center}

\begin{figure}[t]
\plotfiddle{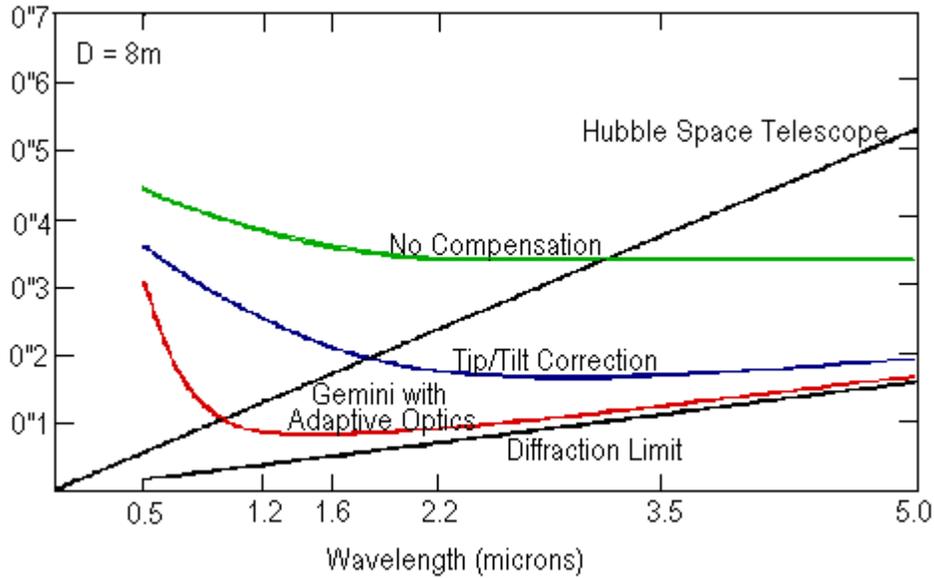}{8cm}{0}{75}{75}{-190}{-10}
\caption{Predicted optimal image quality versus wavelength for HST and an
8-meter telescope with three levels of correction: no correction,
tip/tilt correction, and adaptive optics.  The Gemini telescopes
always use tip/tilt correction and will eventually have facility
adaptive optics systems that can feed any instrument.  Therefore, in
the near-IR the image quality produced by Gemini is equal to or better
than that produced by HST.}
\end{figure}

\section{GMOS}

The Gemini Multi-Object Spectrograph (GMOS) instruments --- one for
each telescope --- provide the primary optical imaging and
spectroscopic capabilities at Gemini (Murowinski et al. 2002).  Each
instrument will have a lenslet/fiber-based IFU that sits in one of the
three mask cassettes and it is deployed like any other GMOS slit mask.
The IFUs are built by the University of Durham and the design of the
first unit is described by Allington-Smith et al. (2002; also see
Allington-Smith et al. in these proceedings).  Commissioning for the
first IFU was begun in September, 2001, and its performance has met or
exceeded expectations. It was first offered to the community for use
in the 2002A semester.

As the design of the IFU is covered elsewhere, this paper will focus
on the data format and data reduction procedure.  The IFU has two
sub-fields --- one with 1000 lenslets ($7''\times5''$), and one with
500 lenslets ($3\farcs5\times5''$) ---
separated by 1 arcminute so that one field can be used for sky
subtraction.  The light entering the lenslets is redirected by fibers
into two pseudo-slits of 750 fibers each.  These are imaged on the
detector like regular longslits.  The current detector is an array of
three butted $2048\times4608$ pixel EEV CCDs with an effective size of
$6144\times4608$.  The gaps between the CCDs are equivalent to 37
pixels in width.  The dispersion axis is along the long axis of the
mosaic. When both IFU slits are used two banks of 750 spectra appear
side-by-side on the detector and blocking filters must be used to avoid
spectral overlap.  However, either of the slits can be closed,
resulting in one-half the field-of-view but allowing twice the
spectral coverage.  A raw GMOS file is a multi-extension FITS (MEF)
file with one extension for each amplifier used for readout (usually 3
or 6).

All Gemini pipeline processing will be done within IRAF.  The Gemini
staff are writing packages of scripts for handling the data from
facility instruments.  Scripts for handling GMOS IFU data will be part
of the GMOS package and will allow for the extraction, calibration,
and analysis of the IFU spectra (Table~2).  Most of these scripts are
in draft form and the first release is planned for the middle of
semester 2002A.

\begin{center}
Table 2. IFU related scripts in the GMOS IRAF package
\begin{tabular}{|l|l|}
\hline
Task & Function \\
\hline
{\sf gfapsum} & sum spectra in a spatial region\\
{\sf gfdisplay} & display datacubes using {\sf ldisplay}\\
{\sf gfextract} & extract spectra, apply fiber throughput correction\\
{\sf gfmosaic} & merge datacubes\\
{\sf gfquick} & quick image reconstruction for target acquisition\\
{\sf gfreduce} & apply reduction/calibration to object frames\\
{\sf gfresponse} & determine relative fiber responses\\
{\sf gfskysub} & subtract sky\\
{\sf gftransform} & apply wavelength calibration\\
{\sf gscrrej} & remove cosmic rays\\
{\sf gsreduce} & bias subtraction \\
{\sf gswavelength} & determine wavelength calibration\\
\hline 
\end{tabular}
\end{center}

\newpage
The current extraction routine uses the IRAF task {\sf apall} to find,
trace, and extract the spectra.  Most of the spectra are separated
well enough that identifying the individual spectra is not a problem.
However, the peaks of three low-throughput fibers cannot be
distinguished from their neighbors, so they are lost.  More
sophisticated reduction techniques, such as deconvolution, may be able to
recover these spectra.  The current format of the extracted
``datacube'' is shown in Figure~2.  The spectra are packed in 2D IRAF
multispec images (one spectrum per image line) within a MEF file.  The
relative positions of the lenslets on the sky are contained in the
binary table (MDF) extension.  This format is similar in concept to
the proposed Euro3D format.  Tasks such as {\sf gfdisplay} are used to
visualize the 3D data.  Additional analysis tools will be released as
they are developed.

\begin{figure}
\plotone{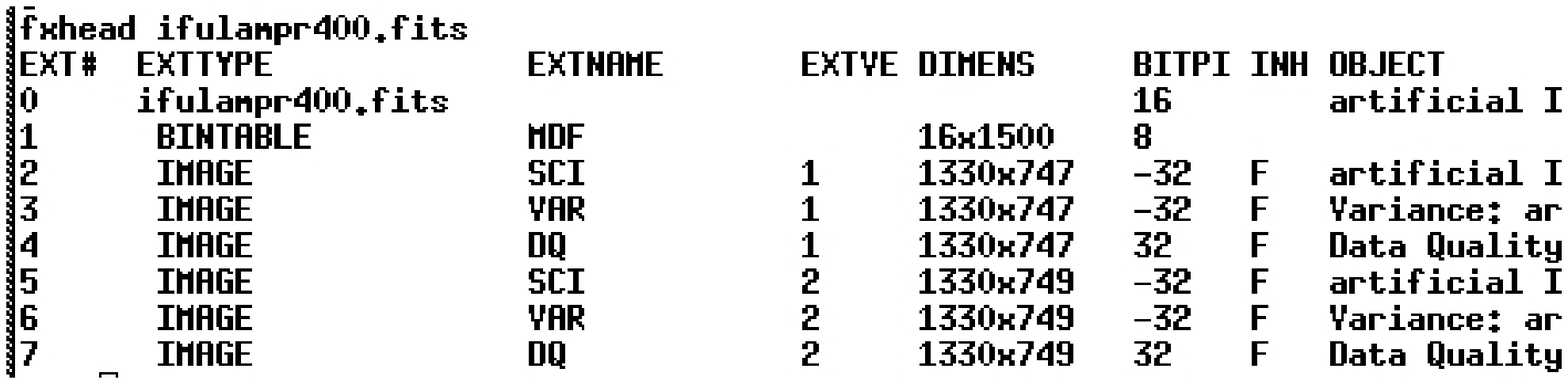}
\caption{The format of an example GMOS IFU extracted datacube.  The
Mask Definition File (MDF) is a binary table that contains information
about each lenslet, such as relative position on the sky.  There can
be up to three extensions for each slit block extracted.  Each image
plane is an IRAF multispec file (one spectrum per line).  The SCI
extension has the extracted spectra.  The optional VAR and DQ planes
hold the variances of the spectra and the data quality flags.}
\end{figure}

\section{GNIRS}

The Gemini Near-Infrared Spectrograph (GNIRS) is a fully cryogenic
instrument sensitive from 1 to 5 microns.  The instrument is being
built by NOAO in Tucson and the IFU module is being built by the
University of Durham.  Similar to the GMOS IFU, the GNIRS IFU is in a
cassette that is inserted into the beam on a slit slide.  Since the
instrument is cooled the IFU is an image slicer containing 21
diamond-turned mirrors that reformat the focal plane into a bank of
spectra (Figure~3). The width of a slicer mirror is $0\farcs15$ and
the slitlets are 4\farcs4 long, giving 609 independent spatial
elements.  With slices of this width the IFU is optimized for tip/tilt
correction rather than adaptive optics. Spectral resolutions for the
short camera are $R=667$, $R=2000$, and $R=6000$, depending on
the grating.  Delivery to Gemini South is expected to be toward the
end of 2002.\\

\begin{figure}
\plotone{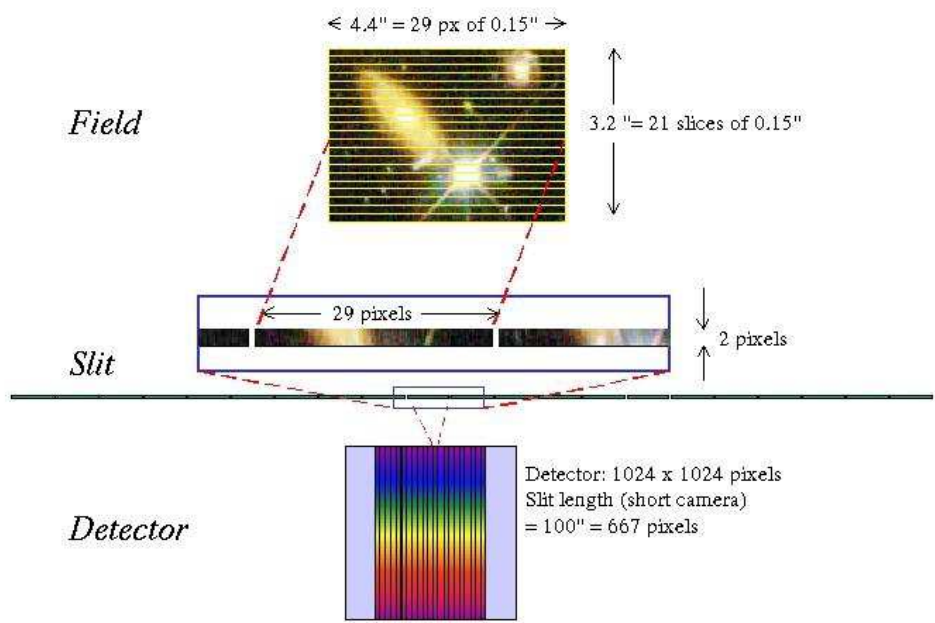}
\caption{Schematic of the image-slicer concept used by GNIRS and
NIFS (Allington-Smith, private communication).  Specifications are for
the GNIRS IFU. Cooled aluminum slicing mirrors divide the focal plane into a
stack of slitlets.  Reduction will be similar to infrared MOS slit spectroscopy
except that full 2D spatial information is preserved.}
\end{figure}
\section{NIFS}

The Gemini Near-infrared Integral Field Spectrograph (NIFS), being
built by the Australia National University, is designed to work behind
the Altair adaptive optics system on Gemini North.  To reduce cost and
speed development the project is copying the the designs of the
on-instrument wavefront sensor and the cryostat from the NIRI
instrument already in use at Gemini North.  The IFU is based on an
image slicer similar to that used in GNIRS.  The slicing mirrors have
a projected width of $0\farcs1$ and the total field-of-view is
$3''\times3''$. The 2048$^2$ Hawaii-II detector will give wavelength
coverage from 0.9--2.5~$\mu$m with a spectral resolution of
5000. Delivery is expected in the middle of 2003.

\section{Future}

Future integral field spectrographs are likely to take advantage of
the multi-conjugate adaptive optics system being developed for Gemini
South.  In this system multiple deformable mirrors will produce a
uniform, near diffraction-limited PSF over a 2 arcminute field.
Several designs for a spectrograph with multiple, deployable IFUs are
under consideration.

\section{Summary}

Large collecting areas, good image quality (with or without adaptive
optics), and infrared optimization make the Gemini telescopes
well-suited for integral field spectroscopy.  Therefore, Gemini will
be offering both optical and near-IR IFU capability at both telescopes
within the next two years.  These instruments will be powerful tools
for studies of galaxy dynamics, black holes, and ISM kinematics
and abundances, to name a few possible projects.

\acknowledgements

The Gemini Observatory is operated by the Association of Universities
for Research in Astronomy, Inc., under a cooperative agreement with
the NSF on behalf of the Gemini partnership: the National Science
Foundation (United States), the Particle Physics and Astronomy
Research Council (United Kingdom), the National Research Council
(Canada), CONICYT (Chile), the Australian Research Council
(Australia), CNPq (Brazil) and CONICET (Argentina).


\begin{references}

\reference Allington-Smith, J., Murray, G., Content, R., Dodsworth,
G., Davies, R., Jorgensen, I., Miller, B. W., Hook, I., Crampton, D.,
\& Murowinski, R. 2002, in preparation

\reference Murowinski, R., et al. 2002, in preparation

\end{references}
\end{document}